\documentclass{nature}

\usepackage{amsmath,amssymb}
\usepackage{xcolor}

\usepackage{caption}

\usepackage{bm}
\usepackage{braket}
\usepackage{amssymb}
\usepackage{textcomp}

\usepackage{bbold}
\usepackage{mathdots}
\usepackage[thicklines]{cancel}

\usepackage{hyperref}
\hypersetup{%
	breaklinks=true,%
	citecolor=blue,%
	colorlinks=true,%
	linkcolor=[RGB]{0,0,200},%
	urlcolor=blue
}	

\usepackage{makecell}
\newcommand{\equa}[1]{Eq.~\ref{#1}}

\renewcommand{\Re}[1]{\mathrm{Re}\left[{#1}\right]}

\title{Manipulating the symmetry of photon-dressed electronic states}

\author{Changhua Bao$^{1,2}$, Michael Sch\"uler$^{3,4}$, Teng Xiao$^{1}$, Fei Wang$^{1,2}$,  Haoyuan Zhong$^{1,2}$, Tianyun Lin$^{1,2}$, Xuanxi Cai$^{1,2}$, Tianshuang Sheng$^{1,2}$, Xiao Tang$^{1,2}$, Hongyun Zhang$^{1,2}$, Pu Yu$^{1,2,5}$, Zhiyuan Sun$^{1,5}$, Wenhui Duan$^{1,2,5,6}$ and Shuyun Zhou$^{1,2,5,\dagger}$}

\usepackage{graphicx}
\makeatletter
\let\saved@includegraphics\includegraphics
\AtBeginDocument{\let\includegraphics\saved@includegraphics}

\makeatother

\begin{document}
\maketitle
\begin{affiliations}
\item Department of Physics, Tsinghua University, Beijing 100084, People's Republic of China
\item State Key Laboratory of Low-Dimensional Quantum Physics, Tsinghua University, Beijing 100084, People's Republic of China
\item Laboratory for Materials Simulations, Paul Scherrer Institute, PSI, Switzerland
\item Department of Physics, University of Fribourg, Fribourg, Switzerland
\item Frontier Science Center for Quantum Information, Beijing 100084, People's Republic of China
\item Institute for Advanced Study, Tsinghua University, Beijing 100084, People’s Republic of China

 $\dagger$e-mail: syzhou@mail.tsinghua.edu.cn
\end{affiliations}


\newpage

\begin{abstract}
{
\bf Abstract\\
Strong light-matter interaction provides opportunities for tailoring the physical properties of quantum materials on the ultrafast timescale by forming photon-dressed electronic states, i.e., Floquet–Bloch states.  While the light field can in principle imprint its symmetry properties onto the photon-dressed electronic states, so far, how to experimentally detect and further engineer the symmetry of photon-dressed electronic states remains elusive. Here by utilizing time- and angle-resolved photoemission spectroscopy (TrARPES) with polarization-dependent study, we directly visualize the parity symmetry of Floquet–Bloch states in black phosphorus.  The photon-dressed sideband exhibits opposite photoemission intensity to the valence band at the $\bm\Gamma$ point, suggesting a switch of the parity induced by the light field. Moreover, a ``hot spot'' with strong intensity confined near $\bm\Gamma$ is observed, indicating a momentum-dependent modulation beyond the parity switch.  Combining with theoretical calculations, we reveal the light-induced engineering of the wave function of the Floquet–Bloch states as a result of the hybridization between the conduction and valence bands with opposite parities, and show that the ``hot spot'' is intrinsically dictated by the symmetry properties of black phosphorus. Our work suggests TrARPES as a direct probe for the parity of the photon-dressed electronic states with energy- and momentum-resolved information, providing an example for engineering the wave function and symmetry of such photon-dressed electronic states via Floquet engineering.}

\end{abstract}

\newpage

\renewcommand{\thefigure}{\textbf{Fig.~\arabic{figure}}}
\setcounter{figure}{0}

\section*{Introduction}
Symmetry lays the cornerstone for a rich variety of fascinating phases of quantum materials\cite{gross1996role}, for example, the time-reversal symmetry breaking is critical for quantum Hall effect\cite{novoselov2005two}, the U(1) symmetry breaking is associated with superconductivity or superfluidity\cite{tilley2019superfluidity}, and the broken inversion symmetry is fundamental for valleytronics\cite{xu2014spin}. On-demand control of the electronic symmetry paves a highly productive route for accessing new matter phases and realizing new  functionalities\cite{basov2017towards,SunNRP2021}. One promising approach for controlling the symmetry of electronic states is to use strong light–matter interaction to form photon-dressed electronic states\cite{Oka2009PhotoHE,Demler2011PhotoHE, Lindner2011natphy}, which would in principle allow to tailor the electronic structure and even the symmetry of the electronic wave functions inside the material on an ultrafast timescale. Over the past decades, much progress has been achieved in light-tailored electronic structure of the light–matter hybrid system\cite{okaRev2019,Lindner2020NRP,Sentef2021RMP,ZhouNRP2021,bloch2022strongly,hubener2021engineering,kobayashi2023floquet}, including the realization of Floquet engineering\cite{Gedik2013Sci,Gedik2016NP,Cavalleri2020NP,Hsieh2021Nature,Zhou2023Nature,ZhouPRL2023} and Floquet–Bloch states\cite{aeschlimann2021survival,Lee2022Nature,Huber2023Nature}, light-tailored valleytronics\cite{kim2014ultrafast,SieValley2015,sie2018large} etc. Moreover, it has been predicted that the light field could lead to momentum-dependent engineering of the wave function or symmetry properties of the electronic states,  turning an ordinary insulator into a Floquet topological insulator\cite{Lindner2011natphy}.

A central task along the pathway of light-field driven symmetry engineering is the characterization and manipulation of the symmetry properties of photon-dressed electronic states, which, however, remains an experimental challenge. So far, the symmetry properties of photon-dressed electronic states are mainly inferred from light-induced transient optical or transport properties. For example, light-induced inversion symmetry breaking deduced from second harmonic generation\cite{yang2019lightwave,sirica2022photocurrent}, the symmetry properties of Floquet sidebands deduced from the selection rules of high harmonic generation\cite{alon1998selection,saito2017observation,nagai2020dynamical}, and time reversal symmetry breaking induced anomalous Hall effect\cite{Cavalleri2020NP}. These optical and transport measurements provide global symmetry information of the materials under investigation. However, directly revealing the symmetry properties of the electronic wave functions of the hybrid system, in particular with energy- and momentum-resolved information, could provide critical information for the momentum-dependent wave function, which is essential in searching for novel photon-dressed electronic states. Here, by utilizing polarization-dependent time- and angle-resolved photoemission spectroscopy (TrARPES), we directly clarify the parity symmetry of photon-dressed states (Floquet–Bloch states) in black phosphorus. We find that the light field modulates the parity and wave function of the Floguet-Bloch states near the  $\Gamma$ point, leading to a momentum-confined ``hot spot'' which is intrinsically dictated by the symmetry properties of black phosphorus.

\section*{Results}

\begin{figure*}[htbp]
	\centering
	\includegraphics[width=16.8cm]{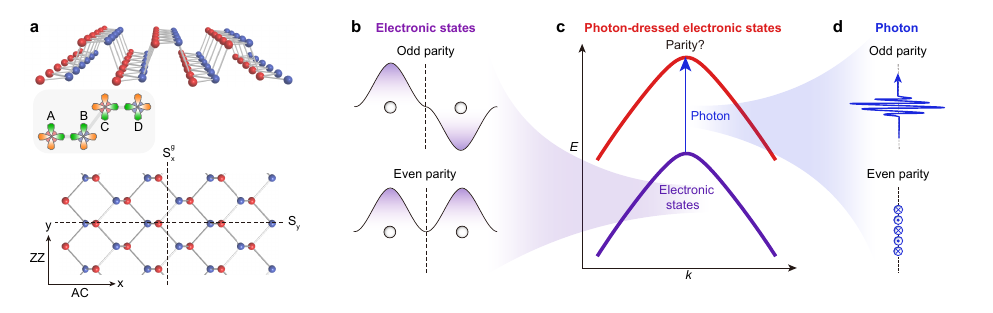}
	\caption{{\bf $\bm{|}$  The symmetry of photon-dressed electronic states.} \textbf{a}, The crystal structure of black phosphorus. The shadows represent the s–p hybridized atomic orbitals with orange and green colors representing different signs. The horizontal and vertical dashed line means the mirror operation $S_y~(y \rightarrow -y)$ and the glide mirror $S^g_x~(x \rightarrow -x, y \rightarrow y+b/2)$ where $b$ is the lattice constant along the ZZ direction. \textbf{b}, Schematic drawings for the wavefunction of electronic states with odd and even parity with respect to the mirror plane (dashed lines). \textbf{c}, Band structure of valence band and photon-dressed first-order Floquet sideband. \textbf{d}, Schematic illustrations for the ultrafast light field with odd and even parity with respect to the mirror plane (dashed lines).
}
\label{fig:schematics} 
\end{figure*}

Black phosphorus exhibits distinctive symmetry operations, which makes it an ideal candidate for investigating the symmetry of photon-dressed electronic states. In black phosphorus, there are two important symmetry operations as schematically illustrated in \ref{fig:schematics}a: the mirror  $S_y: y \rightarrow -y$ which reverses the zigzag (ZZ) direction of the lattice,  and the glide mirror $S^g_x: x \rightarrow -x, y \rightarrow y+b/2$ which reverses the armchair (AC) direction and then translates the lattice in the ZZ direction by half a unit cell (a nonsymmorphic symmetry). The conduction band (CB) has even parity while the valence band (VB) has odd parity\cite{Kim2020natmater,Bao2020natmater} (\ref{fig:schematics}b) under the glide mirror $S^g_x$ which is the focus of this work. Interestingly, the light field can be classified into odd/even parities through the glide mirror $S^g_x$, depending on whether its electric field is parallel or perpendicular to the AC direction, as schematically illustrated in \ref{fig:schematics}d.  The parity properties of the photon-dressed electronic states (\ref{fig:schematics}c) are encoded by both the light field and electrons inside the crystal. In the following, using black phosphorus as a prototypical example, we report the characterization of the parity symmetry of the Floquet–Bloch states under strong light–matter interaction.

\begin{figure*}[htbp] 
	\centering
	\includegraphics[width=16.8cm]{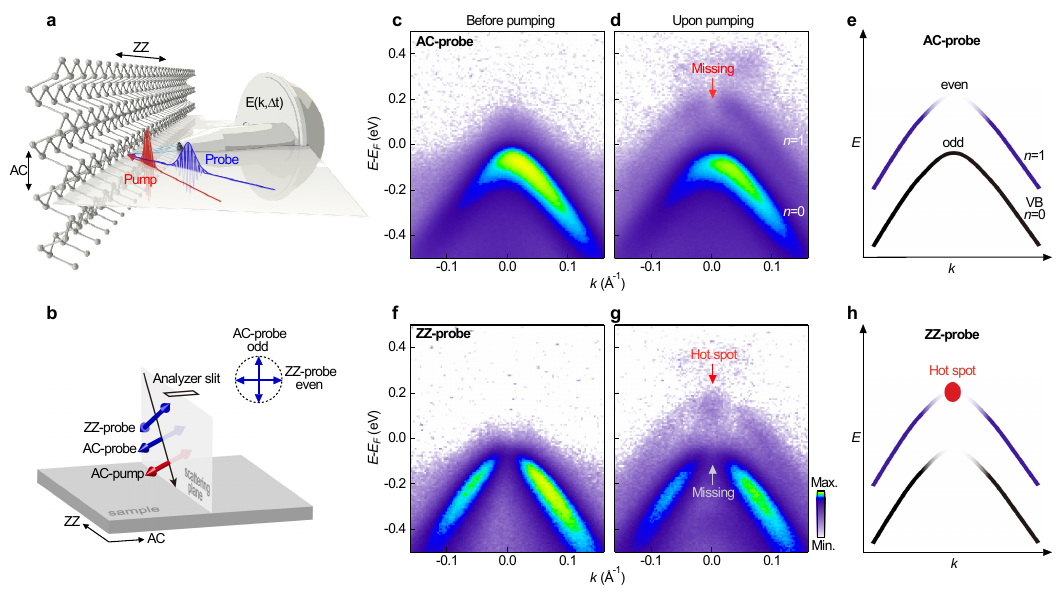}
	\caption{{\bf $\bm{|}$ Observation of the parity modulation and hot spot in the Floquet sideband.} \textbf{a}, A schematic drawing of TrARPES on black phosphorus. \textbf{b}, A schematic for experimental geometry and using different probe polarizations to measure different parities. \textbf{c-e}, TrARPES dispersion images with AC-probe measured along the AC direction ($k$ along the $x$ direction) at $\Delta t$ = -1 ps (\textbf{c}) and $\Delta t$ = 0 (\textbf{d}) and the schematic summary (\textbf{e}) of data in \textbf{d}. \textbf{f-h}, TrARPES dispersion images with ZZ-probe measured along AC direction at $\Delta t$ = -1 ps (\textbf{f}) and $\Delta t$ = 0 (\textbf{g}) and the schematic summary (\textbf{h}) of data in \textbf{g}. The red arrow points to the hot spot. The pump polarization is along the AC direction and perpendicular to the scattering plane, as shown in \textbf{b}. The pump photon energy is 240 meV and the pump fluence is 1.5 mJ/cm$^2$.   
	} 	
\label{fig:spot}
\end{figure*}

To reveal the parity symmetry of Floquet states in black phosphorus, we have performed polarization-dependent TrARPES\cite{ZXRMP21,RevModPhys.96.015003} measurements (\ref{fig:spot}a,b), where the electronic parity is encoded in the polarization of the probe light via the photoemission matrix elements\cite{Kim2020natmater,schuler2020SA,schuler2020PRX}. The photoemission matrix element can be written as $\bra { \phi_f^{\mathbf k}}\mathbf A \cdot \mathbf p \ket{\phi_i^{\mathbf k}}$ (Ref.~\cite{damascelli2003angle}), where $\phi_f^{\mathbf k}$ and $\phi_i^{\mathbf k}$  are final and initial state wavefunctions respectively, $\mathbf A$ is the vector potential of the probe light and $\mathbf p$ is the electron momentum. Since the final-state wavefunction $\phi_f^{\mathbf k}$ is even under reflection with respect to the scattering plane\cite{Kim2020natmater} and the probe light polarized along the AC direction (AC-probe) is odd ($\mathbf A$ is odd),  electrons with odd parity can be probed by AC-probe $\bra { \phi_f^{\mathbf k}}\mathbf A \cdot \mathbf p \ket{\phi_i^{\mathbf k}} = \bra{+} - \ket{-}\neq 0$ while those with even parity cannot be probed by AC-probe $\bra{+} - \ket{+} =  0$ (Ref.~\cite{Kim2020natmater}). For probe light polarized along the ZZ direction (ZZ-probe), the light is even and only electrons with even parity can be probed $\bra{+} + \ket{+}\neq 0$, while those with odd parity cannot be probed by ZZ-probe $\bra{+} +\ket{-} =  0$.  

Using AC-probe which is favorable for detecting the VB with odd parity (\ref{fig:spot}c), we find that upon AC pumping, the first-order Floquet sideband ($n$ = 1) is clearly observed in \ref{fig:spot}d with an energy shift equal to one pump photon energy. Interestingly, the Floquet sideband is not just a simple copy of the VB with $n$ = 0, but instead, there is a strong intensity modulation. In particular, the spectral weight is strongly suppressed at the $\Gamma$ point of the $n$ = 1 sideband, which is in sharp contrast to the VB, where a strong intensity is observed at the $\Gamma$ point, as schematically illustrated in \ref{fig:spot}e. Such strong intensity contrast suggests that the electronic parity is modified by the pump light field from odd to even, since the AC-probe can only have non-zero photoemission matrix elements for electrons with odd parity.

The observed parity switch demonstrates the imprint of the symmetry of light fields on a light–matter hybrid system experimentally. Since an AC-pump light field is applied where the electric field is perpendicular to the scattering plane (\ref{fig:spot}b), it has odd parity under the glide mirror $S_x^g$. According to our experimental observation, the parity of the first-order Floquet sideband at the $\Gamma$ point can be phenomenologically written as a combination of the parities of the pump light and the electron: $\ket{\rm sideband} = \ket{\rm pump, AC} \otimes \ket{\rm electron, VB} = \ket{\rm odd} \otimes \ket{\rm odd} = \ket{\rm even}$. The parity switch is further confirmed by TrARPES measurements using ZZ-probe while keeping the same pump polarization along AC (\ref{fig:spot}f,g).  Again, the edge of the first-order Floquet sideband shows an opposite response to the VB, with a strong intensity for the Floquet sideband at the $\Gamma$ point, in contrast to the  VB. In particular, an isolated ``hot spot'' is observed in the first-order Floquet sideband (indicated by the red arrow). This feature is schematically summarized in \ref{fig:spot}h. It is also observed in different samples, indicating that it is an intrinsic effect (Supplementary Figure 1).

\begin{figure*}[htbp]
	\centering
	\includegraphics[]{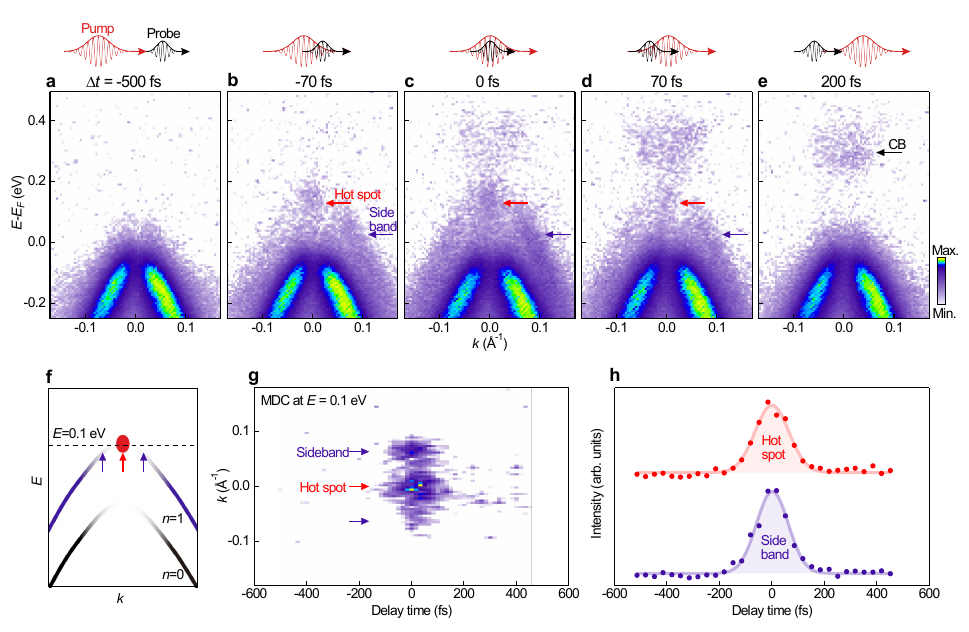}
	\caption{{\bf $\bm{|}$  Co-development of the hot spot and light fields in the time domain.} \textbf{a-e}, TrARPES dispersion images measured at different delay times. The red and purple arrows point to the hot spot and Floquet sideband $n$ = 1, respectively. \textbf{f,g}, Continuous evolution of momentum distribution curves as a function of delay time at $E$ = 0.1 eV (\textbf{g}) as schematically illustrated in panel \textbf{f}. \textbf{h}, Intensity of hot spot and sideband $n$ = 1 as a function of delay time, which is fitted by Gaussian functions.} 
	\label{fig:timedependence}
\end{figure*}

To explore the origin of the hot spot, its evolution in the time domain is revealed. \ref{fig:timedependence} shows snapshots of dispersion images measured at different delay times with ZZ-probe.  It is clear that the hot spot is observed only around time zero (indicated by red arrows in \ref{fig:timedependence}b-d), simultaneously with the Floquet sideband (indicated by purple arrows). At a later delay time when the sideband disappears (\ref{fig:timedependence}e), the hot spot also disappears, leaving only a long-lived CB intensity.  \ref{fig:timedependence}g,h shows continuous evolution of the momentum distribution curve (MDC) at $E$ = 0.1 eV, which cuts through the hot spot and residual intensity of the sideband (\ref{fig:timedependence}f). The data clearly show that the hot spot spans a temporal window of 160 fs determined by the pump and probe pulses and co-developes with the Floquet sideband. The simultaneous development of the hot spot and the Floquet sideband in the time domain suggests that the hot spot is strongly related to the Floquet states.

\begin{figure*}[htbp]
	\centering
	\includegraphics[width=16.8cm]{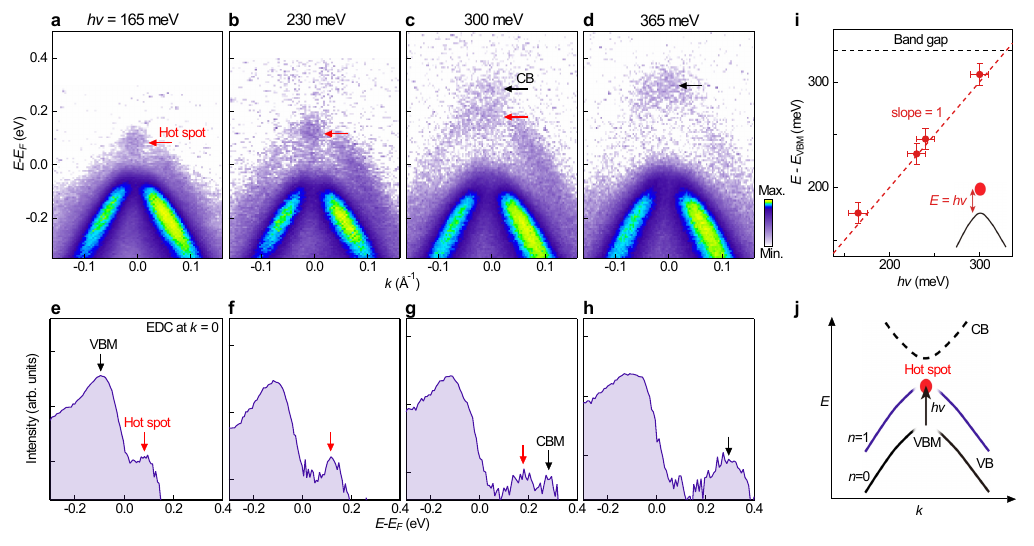}
	\caption{{\bf $\bm{|}$  Pump photon energy dependence of the hot spot.} \textbf{a-d}, TrARPES dispersion images measured at $\Delta t$ = 0 with different pump photon energies. The red and black arrows point to the hot spot and CB, respectively. \textbf{e-h}, Energy distribution curves at $k$ = 0 with different pump photon energies, which are correspondingly extracted from the data in \textbf{a-d}. \textbf{i}, Extracted energy positions of the hot spot relative to the energy of VBM as a function of pump photon energy. The red dashed line corresponds to $E = h\nu$. The error bar of the energy position is defined as the upper limit when the energy position is clearly offset from the peak, and the error bar of the photon energy is defined by the wavelength deviation between the setting and real values of non-collinear differential frequency generation (NDFG). \textbf{j}, A schematic drawing to illustrate that the hot spot is the Floquet replica of VBM.} 
\label{fig:hvdependence}
\end{figure*}

The Floquet sideband nature of the hot spot is further confirmed by the pump photon energy dependent measurements in \ref{fig:hvdependence}a-d. As the pump photon energy increases, the hot spot shifts to higher energy, consistent with the behavior of the Floquet sideband. When approaching the above-gap pumping region (\ref{fig:hvdependence}c,d),  the allowed optical transition between the VB and CB makes it difficult to trace the hot spot. To explore the origin of the hot spot, a quantitative analysis of the energy values for the hot spot and the VB maximum (VBM) is performed from the energy distribution curve (EDC)   at the $\Gamma$ point (\ref{fig:hvdependence}e-h). The extracted values are plotted in \ref{fig:hvdependence}i. It is clear that the energy positions of the hot spot show a linear scaling with the pump photon energy as $E=h\nu$, again confirming that the hot spot is the first-order Floquet sideband, with an energy shift equal to one pump photon energy, as schematically illustrated in \ref{fig:hvdependence}j.

Since the hot spot is the Floquet sideband, an intriguing question is why the light-field dressed Floquet sideband exhibits such a dramatic intensity modulation near the $\Gamma$ point. To answer this question, first-principles calculations and TrARPES simulation with the time-dependent non-equilibrium Green's function approach are performed (see  Methods for more details). First, the hot spot is well reproduced for ZZ-probe case as shown in \ref{fig:theory}a (calculations for all geometries are shown in Supplementary Figure 2, 3), which shows good agreement with the experimental results. To reveal the origin of the hot spot, we have further projected the calculated TrARPES spectral weight onto the VB and CB respectively  (\ref{fig:theory}b,c). The projection clearly shows that the hot spot has a strong contribution from the CB, although it is the light-field dressed sideband of the VB. We further plot the momentum-dependent  CB/VB spectral weight of the VB Floquet eigenstate under different pump fluences in \ref{fig:theory}d.  The spectral weight contribution from the CB orbital is mainly confined around the $\Gamma$ point and can approach $\sim$20$\%$ under an experimentally realistic electric field of 800 kV/cm, which corresponds to a reduced spectral weight contribution in the VB orbital. This suggests that the wave function of the VB Floquet eigenstate around the $\Gamma$ point is strongly modulated to exhibit both characteristics of CB and VB.

\begin{figure*}[htbp]
	\centering
	\includegraphics[]{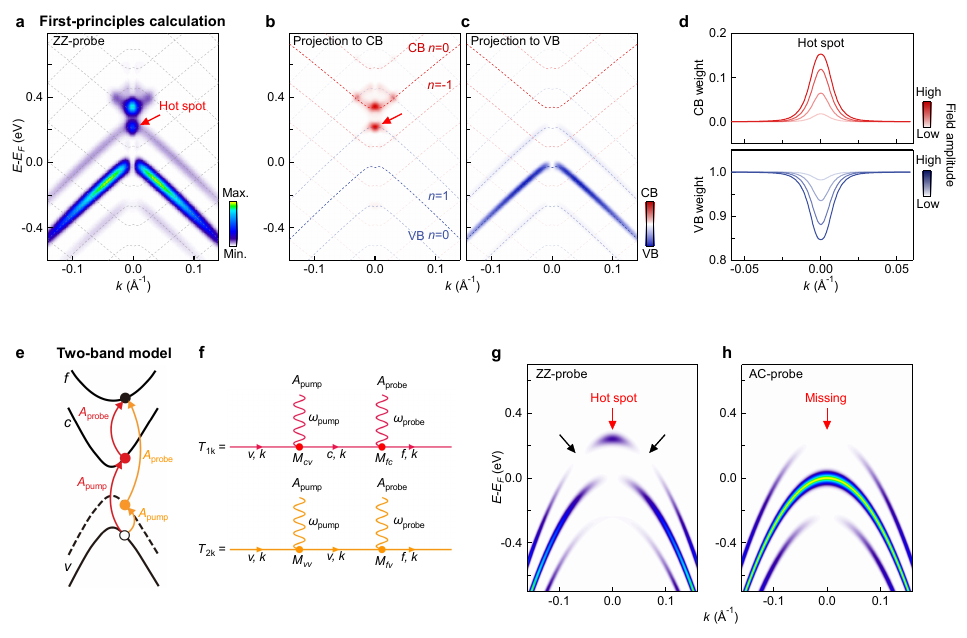}
	\caption{\footnotesize {\bf $\bm{|}$  Theoretical calculations to reveal the origin of the hot spot.} \textbf{a}, Calculated TrARPES spectra upon pumping for ZZ-probe case from first-principles calculation. \textbf{b,c}, Projection of the data in \textbf{a} to CB (\textbf{b}) and VB (\textbf{c}), respectively. The dotted curves are the corresponding dispersions. A field amplitude of $E_0$ = 500~kV/cm is used for the simulations. \textbf{d}, Spectral weight of the VB Floquet eigenstate (fully occupied if electron–electron scattering is neglected) projected to CB and VB as a function of momentum. Field amplitudes from 200 to 800~kV/cm are used with an interval of 200~kV/cm. The measured first-order Floquet sideband may be viewed as the $n= 1$ component of the valence band Floquet eigenstate.  \textbf{e}, Schematic of the the interband transition path $T_{1k}$ (red arrows) and the intraband path $T_{2k}$ (orange arrows) in the two-band model.  In each path, the first step (the pumping process) is a virtual process that should not be confused with a resonant excitation. $A_{\text{probe}}$, $A_{\text{pump}}$, $\omega_{\text{probe}}$ and $\omega_{\text{pump}}$ are the vector potentials and angular frequency of the probe and the pump. $v$, $c$, and $f$ are the valence band, conduction band, and final free-electron state, respectively. $k$ is momentum. $M_{cv}$, $M_{vv}$, $M_{fc}$ and $M_{fv}$ are corresponding transition matrix elements. \textbf{f}, The Feynman diagrams of the two paths in \textbf{e}.  \textbf{g,h}, Simulated TrARPES spectra for ZZ-probe (\textbf{g}) and AC-probe (\textbf{h}), respectively, plotted from \equa{eqn:amplitude} with AC-pump.} 
\label{fig:theory}
\end{figure*}

Such observation is in line with recent reports on Floquet engineering, namely, light-field induced modulation of the transient electronic structure, upon near-resonance and below-gap pumping of black phosphorus\cite{Zhou2023Nature,ZhouPRL2023}.  Here we go one step further to show that not only the electronic band structure is renormalized, but also the wave functions are tailored by the light field. Therefore, the hot spot is a smoking-gun signature of the light-induced modulation of wave functions, and we have shown that such wave function modulation is strongly momentum-dependent.

\section*{Discussion}

To obtain more insights into the microscopic mechanism for the formation of hot spot, an analytical two-band model is considered. As shown in \ref{fig:theory}e,f, two excitation paths are involved: the interband path $T_{1k}$ where the  electron's wave function changes from the VB  orbital to the CB orbital as a virtual process by the pump (with the matrix element $M_{cv}$) before being excited into a free-electron final state by the probe ($M_{fc}$); the intraband path $T_{2k}$ where the electron is dressed  as a virtual process by the pump ($M_{vv}$) without changing its orbital character, and then excited to the free-electron state by the probe ($M_{fv}$). The TrARPES intensities of the  Floquet sidebands  can be written as $I_k^{\pm1} \propto |T_k(\pm \omega_{\text{pump}})|^2$, where  the TrARPES  amplitude is
\begin{align}\label{eqn:amplitude}
T_k(\omega)=T_{1k}(\omega)+T_{2k}(\omega)
=
		{\frac{M_{cv,k} M_{fc,k}  }{\omega-(\varepsilon_{k}^c-\varepsilon_{k}^v)}}
+
		\frac{M_{vv,k} M_{fv,k} }{\omega} 
		\,
\end{align}
to the first order in the pump, with the denominators being the frequency mismatches of the virtual processes. Here $\varepsilon_{k}^v=-{ k^2}/{2 m_v}$ and  $\varepsilon_{k}^c=\Delta+{ k^2}/{2 m_c}$ are the energies of the valence and conduction bands,  and $\Delta$ is the band gap. The calculated TrARPES intensity from \equa{eqn:amplitude} is plotted in \ref{fig:theory}g,h and it well reproduces the observed parity modulation and hot spot. 

The two-band model allows us to further reveal the hot spot as a universal feature dictated by the mirror symmetry $S_y$ and the glide mirror symmetry $S^g_x$.  Since the CB and VB are even under $S_y$ while the CB/VB is even/odd under $S^g_x$, it constrains the interband matrix element to be even which can be written as $M_{cv,k} = a$ for the low-energy effective model, and the intraband matrix element to be odd which can be written as $M_{vv,k} = b\cdot k$, where $a$ and $b$ are constants. Similarly, the photoemission matrix element is constrained by $S^g_x$ to be even for CB ($M_{fc}= c$) and odd for VB ($M_{fv}= d\cdot k$) for ZZ-probe, where $c$ and $d$ are constants. Inserting these matrix elements into \equa{eqn:amplitude}, one obtains 
\begin{align}\label{eqn:t1t2}
T_{1k}=\frac {a\cdot c} {\omega-\Delta- (1/{ m_c}+ 1/{ m_v}){ k^2}/2}, ~T_{2k}=\frac {b\cdot d\cdot k^2} {\omega}.
\end{align}
Moreover, the two mirror symmetries also constrain $M_{fc,k} M_{cv,k}$ and $M_{fv,k} M_{vv,k}$  to have the same sign (see Methods for more details), so $a\cdot c$ and $b\cdot d$ should have the same sign. While $\omega-\Delta<0$ for below-gap pumping,  $T_{1k}$ and  $T_{2k}$ have opposite signs, resulting in destructive interference between the two paths. The magnitude of $T_{1k}$ decreases from a finite value $a\cdot c/(\Delta-\omega)$ to zero and the magnitude of $T_{2k}$ increases from zero to infinity with the momentum $k$, so there must be certain momentum $k$ where $T_{1k}+T_{2k}=0$ and these two paths exactly cancel each other. This gives two zero intensity points surrounding the hot spot, as indicated by the black arrows in \ref{fig:theory}g. Therefore, the existence of the hot spot and the surrounding zero intensity points in the Floquet sideband is universal in that it is a direct consequence of the symmetry properties of the two bands under the two mirror operations, independent on the details of the material. 

The observation of the engineered symmetry properties of the Floquet sideband with a hot spot in black phosphorus provides a nice example for revealing the momentum-dependent modulation of the Floquet wave functions, uncovering an intriguing aspect of Floquet engineering. It shows that the photon–dressed electronic states not only inherit a parity symmetry as a product of the light field and electrons at the $\Gamma$ point, but could also exhibit strong momentum-dependent wave function mixing through Floquet engineering. Our work demonstrates TrARPES as a powerful technique for directly probing and manipulating the momentum- and energy-dependent wave function of photon-dressed electronic states, and provides insights into the search for fascinating nonequilibrium states such as Floquet topological insulators\cite{Lindner2011natphy}, where the dramatic manipulation of the wave function could lead to topologically nontrivial phases.

\section*{Methods}

\subsection{Sample preparation}
~\\
	Black phosphorus single crystals were grown by the chemical vapour transport method. A mixture of red phosphorus lump (Alfa Aesar, 99.999\%), tin grains (Aladdin, $\geqslant$99.5\%), and iodine crystals (Alfa Aesar, 99.9\%) was sealed under vacuum in a silica tube. The tube was heated to 400 \textcelsius{} within 2 hours and maintained at 400 \textcelsius{} for 2 hours, then heated to 600 \textcelsius{} and maintained at  600~\textcelsius{} for 1 day. The tube was slowly cooled to 350 \textcelsius{} from 600 \textcelsius{} at a cooling rate of 10~\textcelsius/hour, and then furnace-cooled to room temperature. Millimeter-size and high-quality black phosphorus single crystals were obtained.

\subsection{TrARPES measurements}
~\\
	TrARPES measurements were performed at Tsinghua University with a regenerative amplifier laser with a wavelength of 800 nm and a repetition rate of 10 kHz. The pulse energy is 1.3 mJ. 80$\%$ of the beam is used to drive the optical parametric amplifier (OPA) following a non-collinear differential frequency generation (NDFG) to generate a strong mid-infrared pump beam. The probe beam with a photon energy of 6.2 eV is generated by a three-step fourth harmonics generation process using BBO crystals. The polarization of the probe beam is adjusted by a 1/2 wave plate.

\subsection{Details on first-principles calculations}
~\\
\noindent\textit{Interface model and effective fields}\\
The light-matter interaction at the material-vacuum interface is an intricate process where, in principle, the details of the spatially dependent dielectric function play a role. To account for the interface qualitatively, we use the Fresnel equations to compute, for the incoming electric field $\mathbf{E}_\mathrm{in}$, both the reflected field $\mathbf{E}_\mathrm{r}$ and the transmitted field $\mathbf{E}_\mathrm{t}$. As discussed in ref.\cite{keunecke_electromagnetic_2020}, Floquet effects are driven by the internal effective field, which we parameterize as
\begin{align}
\label{eq:Epump}
\mathbf{E}_\mathrm{pump} = s \mathbf{E}_\mathrm{in} + (1 - s) \mathbf{E}_\mathrm{t} \ .
\end{align}
Here, $s$ is a scaling factor interpolating between affecting only the surface ($s=1$) and the bulk ($s=0$). Consistent with the experimental results, we fix $s\approx 0.5$. We note that the results are only weekly dependent on $s$. 
Similarly, the effective field dresses the photoelectrons, giving rise to laser-assisted photoemission (LAPE), is a superposition of incoming and reflected light:
\begin{align}
\mathbf{E}_\mathrm{LAPE} = f \left( \mathbf{E}_\mathrm{in} + \mathbf{E}_\mathrm{r} \right) \ .
\end{align}
Here, we fix the scaling factor $f\approx 0.5$, which is in good agreement with Neppl \textit{et al.}\cite{neppl_direct_2015}.
The reflected/transmitted fields are computed assuming a dielectric constant of $\epsilon = 8$, which corresponds to the static value in equilibrium.

\noindent\textit{Floquet bands}\\
We performed density-functional theory (DFT) calculations for bulk black phosphorous, using the non-primitive unit cell with $c$-axis parallel to the experimental out-of-plane direction. All calculations were performed with the \textsc{Quantum Espresso} package\cite{giannozzi_quantum_2009} within the PBE exchange-correlation functional. We used the pseudopotentials from the \textsc{PseudoDojo} project\cite{van_setten_pseudodojo_2018} and converged the calculations on a $12\times 12\times 8$ Monkhorst-Pack grid. Next, we constructed a Wannier Hamiltonian including the $s$ and $p$ orbitals using the \textsc{Wannier90} code\cite{pizzi_wannier90_2020}. We chose projective Wannier functions without applying the localization procedure. A rigid-shift scissor operator was applied to the conduction bands, adjusting the band gap to the experimentally observed value of $\Delta=330$~meV.
From the Wannier Hamiltonian, we computed the velocity matrix elements $\mathbf{v}_{\alpha\alpha^\prime}(\mathbf{k})$ following refs\cite{yates_spectral_2007,schuler_gauge_2021}. This procedure allows to describe the light–matter coupling in the velocity gauge:
\begin{align}
	\label{eq:wann_velo}
	H_{\alpha\alpha^\prime}(\mathbf{k},t) = \varepsilon_\alpha(\mathbf{k}) \delta_{\alpha\alpha^\prime} - \mathbf{A}_\mathrm{pump}(t)\cdot \mathbf{v}_{\alpha\alpha^\prime}(\mathbf{k}) + \frac{1}{2}\mathbf{A}_\mathrm{pump}(t)^2 \ ,
\end{align}
where $\varepsilon_\alpha(\mathbf{k})$ are the band energies and where $\mathbf{A}_\mathrm{pump}(t) = 1/\omega_\mathrm{pump} \mathrm{Re}[\mathbf{E}_\mathrm{pump} e^{-i\omega_\mathrm{pump} t} ]$ is the vector potential. From the time-dependent Hamiltonian~\eqref{eq:wann_velo} we constructed the Floquet Hamiltonian
\begin{align}
	\label{eq:floq_ham}
	\mathcal{H}_{n\alpha, n^\prime \alpha^\prime}(\mathbf{k}) = \frac{1}{T_\mathrm{p}}\int^{T_\mathrm{p}}_0 dt\, e^{-i (n-n^\prime)\omega_\mathrm{pump} t} H_{\alpha\alpha^\prime}(\mathbf{k},t) - n \omega_\mathrm{pump} \delta_{n,n^\prime} \delta_{\alpha\alpha^\prime} \ ,
\end{align}
where $T_\mathrm{p} = 2\pi/\omega_\mathrm{pump}$. Diagonalizing Floquet Hamiltonian~\eqref{eq:floq_ham} then yields the Floquet quasienergies $\tilde{\varepsilon}_\lambda(\mathbf{k})$ and the corresponding eigenvectors $F^\lambda_{n,\alpha}(\mathbf{k})$ for the Floquet eigenstates
\begin{align}\label{eqn:floquet_state}
|\phi_{\lambda\mathbf{k}}\rangle_t
 = \sum_{\alpha,n} e^{-i(\tilde{\varepsilon}_\lambda+n\omega_{\text{pump}} )t} F^{\lambda}_{n,\alpha}(\mathbf{k}) |\alpha, \mathbf{k} \rangle . 
\end{align}
Here, $|\alpha, \mathbf{k} \rangle$ is the equilibrium Bloch state of band $\alpha$. Hence, the eigenvectors $F^\lambda_{n,\alpha}(\mathbf{k})$ allows us to directly define the projection onto specific bands: $w_\alpha(\mathbf{k}) = \sum_n |F^\lambda_{n,\alpha}(\mathbf{k})|^2$ for a specific Floquet band $\lambda$. Note that if the pump is adiabatically turned on and electron–electron scattering is neglected\cite{schuler2020PRX},  the valence band Floquet eigenstate ($\lambda=v$) is fully occupied.

We analyze the wave function of the Floquet states in more detail in Supplementary Figure 4a-c. For weak field strength (Supplementary Figure 4a), the band character is almost identical to the pure VB or CB, while for stronger fields significant hybridization of the VB and CB at $k_\mathrm{AC} = 0$. In particular, the first sideband of the VB (VB +1) and the CB hybridize strongly, giving rise to a Mexican-hat-like down-bending of the VB +1 and mixing of the orbital character. The increasing gap between VB +1 and the CB is another signature of stronger hybridization, as shown in Supplementary Figure 4d-e. For strong pump field strength, considerable band hybridization of $\sim 20\%$ can be achieved.

\noindent\textit{TrARPES simulations}\\
For the simulation of TrARPES spectra, we replace the pump vector potential in Eq.~\eqref{eq:wann_velo} by the Gaussian pulse
\begin{align}
    \label{eq:avect_pump}
    \mathbf{A}_\mathrm{pump}(t) = \frac{1}{\omega_\mathrm{pump}} S(t)\mathrm{Re}\left[\mathbf{E}_\mathrm{pump} e^{-i\omega_\mathrm{pump} t}\right] \ ,
\end{align}
where $S(t)$ is the Gaussian envelope function with FWHM chosen as in the experiments. We computed the time-dependent density matrix $\rho_{\alpha\alpha^\prime}(\mathbf{k},t)$ from the time-dependent Hamiltonian~\eqref{eq:wann_velo}, assuming unitary time evolution. We included six bands in the calculation (four valence + two conduction bands). From $\rho_{\alpha\alpha^\prime}(\mathbf{k},t)$, we computed the TrARPES spectra using the time-dependent non-equilibrium Green's function (td-NEGF) formalism. As in Refs.\cite{schuler2020PRX,schuler_theory_2021}, we simplified the formalism by using the generalized Kadanoff-Baym ansatz (GKBA). Within the GKBA, the lesser Green's function is obtained from its equation of motion
\begin{align}
    [i \partial_t - \mathbf{H}(\mathbf{k},t)] \mathbf{G}^<(\mathbf{k},t,t^\prime) = 0 \ ,
\end{align}
where we have employed a compact matrix notation. The density matrix enters through $\mathbf{G}^<(\mathbf{k},t,t) = i \boldsymbol{\rho}(\mathbf{k},t)$. From the Green's function, we computed the TrAPPES intensity as
\begin{align}
    \label{eq:trarpes}
    I(\mathbf{k}_\parallel,\tau) &\propto \mathrm{Im}\sum_{k_z}L(k_z)\sum_{\alpha\alpha^\prime} M^*_{f,\alpha}(\mathbf{k})M_{f,\alpha^\prime}(\mathbf{k})\int^\infty_0 \! dt \int^t_0 \! dt^\prime s(t,\tau) s(t^\prime,\tau) e^{-i \varphi(\mathbf{k},t,t^\prime)} G^<_{\alpha^\prime\alpha}(\mathbf{k},t^\prime, t) \ .
\end{align}
Here, $\mathbf{k}_\parallel$ is the measured in-plane momentum, $\tau$ is the pump–probe delay, which enters the envelope functions $s(t,\tau)$ representing the probe pulse (taken as Gaussian functions). The photoemission matrix elements are denoted by $M_{f,\alpha}(\mathbf{k})$. The phase $\varphi(\mathbf{k},t,t^\prime)$ is defined as
\begin{align}
	\label{eq:phase_lape}
    \varphi(\mathbf{k},t,t^\prime) = \int^t_{t^\prime}d \bar{t} \left[\varepsilon_f(\bar{t}) - \omega_\mathrm{probe}\right] \ ,
\end{align}
where $\omega_\mathrm{probe}$ is the photon energy of the probe pulse and $\varepsilon_f(\bar{t})$ is the light-dressed final state energy. The phase~\eqref{eq:phase_lape} incorporates LAPE. We also have the option to include $k_z$ broadening due to the finite mean-free path of the photoelectrons through the Lorentzian function $L(k_z)$ (centered at $k_z=0$).

The photoemission matrix elements are computed within the Wannier-ARPES approach\cite{day_computational_2019,beaulieu_unveiling_2021}.  In essence, from the Wannier Hamiltonian, we obtain the Wannier presentation of the Bloch states:
\begin{align}
	| \alpha, k \rangle = \frac{1}{\sqrt{N}} \sum_{\mathbf{R},j}e^{-i \mathbf{k}\cdot\mathbf{R}} C_{j\alpha}(\mathbf{k}) \phi_j(\mathbf{r} - \mathbf{r}_j - \mathbf{R}) \ .
\end{align}
Here, $\mathbf{R}$ denotes all lattice sites in the supercell with $N$ repetitions of the unit cell, $\phi_j(\mathbf{r})$ are the Wannier orbitals centered at position $\mathbf{r}_j$ within the unit cell, and $C_{j\alpha}(\mathbf{k}$) is the transformation between orbital and band basis. Working within the dipole gauge and approximating the final states as plane waves, the photoemission matrix elements are approximated by
\begin{align}
	M_{f,\alpha}(\mathbf{k}) = \sum_j e^{-i \mathbf{k}\cdot\mathbf{r}_j} C_{j\alpha}(\mathbf{k})M^\mathrm{orb}_j (\mathbf{k}) \ .
\end{align}
The orbital matrix elements are defined by
\begin{align}
	\label{eq:orbital_mel}
	M^\mathrm{orb}_j (\mathbf{k}) = \int d\mathbf{r}\, e^{-i\mathbf{k}\cdot\mathbf{r}} \mathbf{u}\cdot\mathbf{r}\phi_j(\mathbf{r}) \ ,
\end{align}
where $\mathbf{u}$ is the polarization of the probe pulse. In practice, we evaluate the orbital matrix elements~\eqref{eq:orbital_mel} by assuming $\phi_j(\mathbf{r})\approx R_j(r)Y_{\ell_j m_j}(\Omega_\mathrm{r})$ ($Y_{\ell m}(\Omega_\mathrm{r})$ denote the spherical harmonics). The radial dependence $R_j(r)$ is taken from the pseudopotential.

\subsection{Analytical theory of a two-band model}
~\\
\noindent\textit{The two-band model}\\
The tight binding Hamiltonian of the black phosphorus monolayer  (phosphorene) could be written in the basis of the four orthogonal atomic orbitals $\phi_{A/B/C/D}$ in \ref{fig:schematics}a. For the lowest conduction and valence bands, its is enough to use two orbitals:  $\psi_1=\left(\phi_B+\phi_D\right)/\sqrt{2}$ and  $\psi_2=\left(\phi_A+\phi_C\right)/\sqrt{2}$. In this basis,
the Hamiltonian expanded around the $\Gamma$ point to $O(k^2)$ reads\cite{Pereira2015}:
\begin{equation}\label{Htb}
	\hat{H}_{TB}=\left(\begin{array}{cc}
		\eta_x k_x^2+\eta_y k_y^2 & \Delta/2+\gamma_x k_x^2+\gamma_y k_y^2+\mathrm{i} \chi k_x \\
		\Delta/2+\gamma_x k_x^2+\gamma_y k_y^2-\mathrm{i} \chi k_x & \eta_x k_x^2+\eta_y k_y^2
	\end{array}\right).
\end{equation}
Here, the $y$ ($x$) axis is along the ZZ (AC) direction.
The eigenstates of $\hat{H}_{TB}$ are:
\begin{equation}\label{eigen}
|c/v,k\rangle=\frac{e^{i\theta_k}}{\sqrt{2}}
\left(\pm F(k),\,1 \right),
\quad 
	F(k)= \frac{\Delta/2+\gamma_x k_x^2+\gamma_y k_y^2+\mathrm{i} \chi k_x } {\sqrt{\left(\Delta/2+\gamma_x k_x^2+\gamma_y k_y^2\right)^2+\chi^2 k_x^2}}.
\end{equation}
where $+/-$ corresponds to the conduction/valance band, and $\theta_k$ is introduced for convenience for symmetry considerations.
In the band-diagonalized basis, the Hamiltonian $H=	H_0 + H_{\text{pump}}+ H_{\text{probe}}$ including the  vacuum free electrons  and the relevant terms due to the pump and probe fields is 
\begin{equation}\label{eqn:Hband}
	\begin{aligned}
		\hat{H}_0&=\sum_{{k}}\varepsilon_{k}^v \hat{v}_{k}^{\dagger} \hat{v}_{k}+\varepsilon_{k}^c \hat{c}_{k}^{\dagger} \hat{c}_{k} 
	 +  \varepsilon_k^{f} \hat{f}_{k}^{\dagger} \hat{f}_{k}		
	 ,\\
		\hat{H}_{\text{pump}}&=A_{\text{pump}}(t) \sum_{{k}} 
		\left[
		M_{cc}(k) \hat{c}_{k}^{\dagger} \hat{c}_{{k}} + 
		M_{vv}(k) \hat{v}_{k}^{\dagger} \hat{v}_{{k}} +
		\left(M_{cv}(k) \hat{c}_{k}^{\dagger} \hat{v}_{{k}}+  \text{H.c.} \right)
		\right]
		\\
		&=A_{\text{pump}}(t)  \sum_{k} \Psi^\dagger_k H_{\text{pump}}(k)  \Psi_k
		\\
		\hat{H}_{\text{probe}}&=A_{\text{probe}}(t) \sum_{{k}} 
\left[
M_{fc}(k) \hat{f}_{k}^{\dagger} \hat{c}_{{k}} + 
M_{fv}(k) \hat{f}_{k}^{\dagger} \hat{v}_{{k}} + \text{H.c.} 
\right]
		\\
&=A_{\text{probe}}(t)  \sum_{k} \Psi^\dagger_k H_{\text{probe}}(k)  \Psi_k
	\end{aligned}
\end{equation}
where  $\varepsilon_{k}^v=-{ k^2}/{2 m_v}$, $\varepsilon_{k}^c=\Delta+{ k^2}/{2 m_c}$ ($m_{c/v}={1}/\left[{2\left(\pm\eta_x + \gamma_x + \chi^2 / \Delta\right)}\right]$ for $k \parallel x$), $\varepsilon_{{k}}^{f}$ are the kinetic energies of 
 the valance band, conduction band, free electrons, and $\Psi^\dagger_k= (f_k^\dagger,\,  c_k^\dagger,\, v_k^\dagger)$. For notational simplicity, we set the Planck constant $\hbar$, the elementary charge $e$, and the speed of light to be $1$.   
The vector potential of the pump (probe) is $A_{\text{pump}}(t) ={A}_{\text{pump}}s_1(t)\mathrm{e}^{-\mathrm{i}\omega_{\text{pump}} t}+c.c.$ ($A_{\text{probe}}(t) ={A}_{\text{probe}}s_2(t)\mathrm{e}^{-\mathrm{i}\omega_{\text{probe}} t }+c.c.$), where $s(t)$ is the envelop function.
The matrix elements for the linear coupling terms to the pump are simply obtained from the gauge invariant minimal coupling: $M_{mn}=\langle m, k| \partial_k H_{TB} |n, k \rangle $ where $m, n \in (c,v)$.
The information of the linear coupling terms to the probe is not contained in the tight binding model \equa{Htb}, but will be computed later.
We note that in the case of Zigzag (ZZ) probe (${A}_{\text{probe}} \parallel y$), the nonzero component  $q_y$ of the wave vector of the probe field is essential for nonzero ARPES matrix elements $M_{fc}$ and $M_{fv}$. For notational simplicity, we write the probe field as a spatially uniform one unless when computing  $M_{fc}$ and $M_{fv}$.

We now discuss the symmetry constraints on the pump and probe matrix elements, focusing on the line of momenta $(k_x, k_y)=(k, 0)$. 
The important symmetry operations are the mirror $S_y: y \rightarrow -y$ and the glide mirror $S^g_x: x \rightarrow -x, y \rightarrow y+b/2$ shown in \ref{fig:schematics}b. 
Note that all the atomic orbitals $\phi_{A/B/C/D}$ are invariant under $y \rightarrow -y$ plus appropriate translations, and map to each other under $x \rightarrow -x$ plus appropriate translations.
Combined with the crystal structure, it is easy to see that the two orbitals $\psi_{1/2}$ are invariant under $S_y$  and are exchanged under  $S^g_x$, meaning $\hat{S}_y \psi_{1/2} = \psi_{1/2},\, \hat{S}^g_x \psi_{1/2} = \psi_{2/1} $. Therefore, the conduction and valence bands in  \equa{eigen} are even under $S_y$, meaning 
\begin{align}\label{eqn:Sy}
\hat{S}_y |c/v, k\rangle = |c/v, k \rangle \,.
\end{align}
Upon $S^g_x$, the conduction/valence band is even/odd at the gamma point: $\hat{S}^g_x |c, 0\rangle = |c, 0\rangle ,\, \hat{S}^g_x |v, 0\rangle  = -|v, 0\rangle $. Away from the gamma point, one may always find a gauge $\theta_k$ in  \equa{eigen}  such that
\begin{align}\label{eqn:Sgx}
\hat{S}^g_x |c, k\rangle = |c, -k\rangle ,\quad 
\hat{S}^g_x |v, k\rangle  =-|v, -k\rangle  \,,
\end{align}
which we use as the band basis in \equa{eqn:Hband}.

For the pump field in the ZZ direction ($A_{\text{pump}} \parallel y$), since we discuss the case $k_y=0$,  \equa{eqn:Sy} leads to	$\langle n,k| \hat{j}_y |m,k\rangle = - \langle n,k| \hat{S}_y^\dagger \hat{j}_y \hat{S}_y  |m,k\rangle = - \langle n,k|  \hat{j}_y  |m,k\rangle =0 $ where $m,n \in (c,v)$, $\hat{j}_y$ is the current operator. Therefore, one has $M_{cv}=M_{cc}=M_{vv}=0$. 

For the pump field in the AC direction ($A_{\text{pump}} \parallel x$), since the conduction and valence band have different parities under $S^g_x$, one has the nonzero interband matrix element $M_{cv, k} \approx i\chi$ which is obtained from \equa{eigen}. The intraband matrix elements are simply $M_{cc}=k/m_c,\, M_{vv}=-k/m_v$.

The probe matrix elements should be zero if the probe field is in the ZZ direction ($A_{\text{probe}} \parallel y$) because $\langle n,k| \hat{j}_y |m,k\rangle = - \langle n,k| \hat{S}_y^\dagger \hat{j}_y \hat{S}_y  |m,k\rangle = - \langle n,k|  \hat{j}_y  |m,k\rangle =0 $, where $m,n$ may take any band index include vacuum electrons.  However, if  the incident plane is the $z-y$ plane, the nonzero in-plane momentum along the $y$ direction means that the vector potential is not uniform: $A_{\text{probe}}\exp\left[{-\mathrm{i}\left(\omega_{\text{probe}}t-q_yy\right)}\right]+\text{c.c.}$. Linear expansion in $q_y$ gives a nonzero matrix element (quadrupole) of the current operator between a local atomic orbital and the vacuum electron: $\langle f | q_y \hat{y} \hat{j}_y |n\rangle \equiv b_1^y/2$ where $n \in (A, B, C, D)$.  
This renders $M_{fc}=\langle f|  q_y \hat{y} \hat{j}_y | c, k \rangle \approx b_1^y$ and $M_{fv}=\langle f| q_y \hat{y} \hat{j}_y | v, k \rangle \approx -b_1^y  i\chi k/G$. This scaling could be obtained by the symmetry under $S^g_x$ (\equa{eqn:Sgx}):  $M_{fm}(k)=\langle f,k| H_{\text{probe}} |m,k\rangle = \langle f,k| \hat{S}^{g\dagger}_x H_{\text{probe}}  \hat{S}^g_x |m,k\rangle = \pm \langle f,-k| H_{\text{probe}} |m,-k\rangle =\pm M_{fm}(-k)$ where $+/-$ corresponds to $m=c/v$. If the incident plane is the $z-x$ plane, symmetry under $S_y$ (\equa{eqn:Sy}) means $M_{fc}=M_{fv}=0$.
The matrix elements for all the probe configurations are summarized in Supplementary Table 1.

\noindent\textit{Relating the ARPES intensity to Green's functions}\\
The time-accumulated ARPES intensity at in-plane momentum $k$ is 
\begin{equation}
	\begin{aligned}
		I_k^{\text{sum}}=\int_{t_0}^{t_f} \text{d} t \frac{\partial}{\partial t} \rho_{k}^{f}
		=\int_{t_0}^{t_{f}} \text{d} t I_{k}(t) ,
	\end{aligned}
\end{equation}
where $I_{k}(t)$ is the ARPES intensity, $\rho_{k}^{f}=\left\langle\hat{f}_{k}^{\dagger} \hat{f}_{k}\right\rangle =-\mathrm{i}G^<_{ff}(t,t)$, and $G^<_{ff}$ is the lesser Green's function of the vacuum electrons collected by the detector.
To eliminate the time derivative, we use the Heisenberg equation of motion
$i \partial_t \hat{A}=[\hat{A}, \hat{H}] $ :
\begin{equation}
\begin{aligned}
		I_{k}(t)=&\frac{\partial}{\partial t} \rho_{k}^{f}=-\mathrm{i}\frac{\partial}{\partial t}G^<_{ff}(t,t)
		= 2\Re{\sum_{m=\{c,v\}}  M_{mf}(k)A_{\text{probe}}(t)G^<_{fm}(t,t)},
\label{eqn:Ik_Gfm}
\end{aligned}
\end{equation}
where $G^<_{fm}(t,t^{\prime})=\mathrm{i}\left\langle  \hat{m}^{\dagger}_k(t^{\prime}) \hat{f}_k(t)\right\rangle$.

From \equa{eqn:Ik_Gfm} and taking the envelop function $s(t)=1$,  the ARPES intensity is related to the lesser Green's function of electrons inside the material\cite{Stefanucci_vanLeeuwen_2013,Haug_Jauho_2008}:
\begin{equation}\label{eqn:I_Glesser}
	\begin{aligned}
I_{k}(t)
		=&-\mathrm{i} A_{\text{probe}}^2 \sum_{m,n}M_{mf}M_{fn}G^<_{nm}(\epsilon_k^f\pm\omega_{\text{probe}}).
	\end{aligned}
\end{equation}
see \equa{eq:trarpes}.
Here $G^<$  is the dressed Green's function that contains the effect of the pump. Note that on the right hand side of \equa{eqn:I_Glesser}, there is  no integral over $t$ although the result does not depend on it for $s(t)=1$.

\noindent\textit{TrARPES intensity}\\
In TrARPES experiments, the pump field dresses the electrons inside the material, modifying its Green's function. The dressed Green's function $G^<_{nm}$ could be computed to the leading order of the pump as
\begin{equation}\label{eqn:Glesser}
	\begin{aligned}
		G^<(t,t')=&G_0^<(t,t')+\int^{+\infty}_{-\infty}\text{d}{t_1}\int^{+\infty}_{-\infty}\text{d}{t_2}G^r_{0}(t,t_1)H(t_1)G^<_{0}(t_1,t_2)H(t_2)G^a_{0}(t_2,t')
	\end{aligned}
\end{equation}
where $H(t)=H_{\text{pump}}(k)A_{\text{pump}}(t)$ and $G^r_{0}$ ($G^a_{0}$) is the retarded (advanced) Green's function.
After some algebra and plugging  into \equa{eqn:I_Glesser} the $G^<$ from \equa{eqn:Glesser}, the TrARPES intensity  is obtained as $I_{k}=I^v_{k}+I^{+1}_{k}+I^{-1}_{k}$, where $I^v_{k}$ and $I^{+1}_{k}$, $I^{-1}_{k}$ are the intensities of the valence band and its replica of index $+1$,  $-1$. Their expressions are 
\begin{equation} \label{eqn:I}
	\begin{aligned}
		I^v_{k}
		=& {2\pi}\delta\left( {\varepsilon_{{k}}^f}-\omega_{\text{probe}}-\varepsilon_{k}^v\right)  \left( M_{fv}\right) ^2 A_{\text{probe}}^2,\\
		I^{+1}_{k}=
		&2{\pi}\delta\left( {\varepsilon_{{k}}^f}-\omega_{\text{probe}}-\omega_{\text{pump}}-\varepsilon_{k}^v\right) \left| T(\omega_{\text{pump}})\right|^2 A_{\text{probe}}^2A_{\text{pump}}^2,\\
		I^{-1}_{k}
		=&2{\pi}\delta\left( {\varepsilon_{{k}}^f}-\omega_{\text{probe}}+\omega_{\text{pump}}-\varepsilon_{k}^v\right) 
		\left| T(-\omega_{\text{pump}})\right|^2 A_{\text{probe}}^2A_{\text{pump}}^2,
	\end{aligned}
\end{equation}
where the ARPES amplitude $T_k=T_{1k}+T_{2k}$ for the first order replica bands has contributions from two paths: 
\begin{equation}\label{eqn:T}
	\begin{aligned}
		T_{1k}(\omega_{\text{pump}})=& 
		{\frac{1}{\omega_{\text{pump}}-(\varepsilon_{k}^c-\varepsilon_{k}^v-\mathrm{i}0^+)}}
		M_{fc}  M_{cv},
		\quad
		T_{2k}(\omega_{\text{pump}})=
		\frac{1}{\omega_{\text{pump}}} M_{fv} M_{vv}
		\,.
	\end{aligned}
\end{equation}

One may now make connections to the Floquet eigenstates. $T_{k}(\omega_{\text{pump}})$ is the ARPES amplitude of the $n=1$ sideband of the VB, in other words, the $n=1$ component of the VB Floquet eigensate $|\phi_{v\mathbf{k}}\rangle_t = \sum_{\alpha,n} e^{-i(\tilde{\varepsilon}_\lambda+n\omega_{\text{pump}} )t} F^{v}_{n,\alpha}(\mathbf{k}) |\alpha, \mathbf{k} \rangle$, see \equa{eqn:floquet_state}. This Floquet eigensate is fully occupied because if scattering is neglected and the pump is slowly turned on, the VB adiabatically evolves into it.
The coefficients $F_{c,k}\propto F_{1,c}^v (k)$ and $F_{v,k}\propto F_{1,v}^v (k)$  in \equa{eqn:T} are just the superposition coefficients of the $n=1$ sideband in terms of the CB and VB wave functions to the first order of the pump. 
Plugging in the matrix elements from the two-band model, one obtains $F_{c,k}=i\chi/(\omega_{\text{pump}}-(\varepsilon_{k}^c-\varepsilon_{k}^v))$ and $F_{v,k}=-k/(m_v \omega_{\text{pump}})$. Therefore, to the first order of the pump, the  $n=1$ sideband of the VB is of pure CB orbital character, leading to the parity switch by the pump and the formation of the hot spot.
Further plugging in the photoemission matrix elements, one obtains
$T_k(\omega)
= i\chi b_1^y \left(
\frac{1}{\omega-(\varepsilon_{k}^c-\varepsilon_{k}^v)}
+
\frac{k^2/( m_v \Delta)}{\omega} 
\right)
$. For below-gap pumping ($\omega=\omega_{\text{pump}}<\Delta$) relevant to $I_k^{+1} $, the energy mismatch between these two paths gives rise to opposite signs, resulting in destructive interference between the two paths. At certain momenta ($k^2 = m \omega_{\text{pump}}$ if $m_c=m_v=m$) set by $T_k=0$, these two paths must exactly cancel each other, giving zero intensity in $I_k^{+1}$ surrounding the two sides of the hot spot.

For an ultrafast probe pulse with nonzero spectra width  $s_2(t)=\exp {\left(-\frac{\left( t-t_{2}\right) ^2}{2\sigma_2^2} \right)}$, its leading effect is to broaden the delta functions for energy conservation in $I_k^{\text{sum}}$ to Gaussians, e.g.,  $\sigma_2^2 \mathrm{e}^{-\sigma_2^2\left({\varepsilon_{{k}}^f}-\omega_{\text{probe}}-\omega_{\text{pump}}-\varepsilon_{k}^v\right)^2}$. \ref{fig:theory}g,h are plotted from Eqs.~\eqref{eqn:I}\eqref{eqn:T} with parameters: $\Delta=0.33~{\rm{eV}}, \chi=3~{\rm{eV\AA}}, \eta_x=1~{\rm{eV\AA^2}}, \gamma_x=4~{\rm{eV\AA^2}}, \omega_{\text{pump}}=0.24~{\rm{eV}}, \sigma_2=0.03~{\rm{eV}}$.

\noindent\textit{The three-dimensional two-band model}\\
The two-band model developed for monolayer black phosphorus (phosphorene, the 2D model) can be extended to multilayer and bulk black phosphorus without qualitative changes. This is because of the weak interlayer tunneling and the simple stacking structure that does not change the symmetry. As a result, the electronic energy–momentum dispersion of bulk black phosphorus is quasi-two dimensional, with a weak dispersion along the out-of-plane momentum $k_z$ \cite{JiaBPARPESPRB2014,JiNC2014,Margot2023NL}.
ARPES selects a $k_z$ with a small uncertainty\cite{damascelli2003angle}, so that  the pump matrix elements are the same as the 2D model for the in-plane polarization of the electric field in our experiment, meaning that the Floquet physics remains the same.
From the measured band gap, our experiment selected the band around the Z point in the Brillouin zone where the  band gap is minimized\cite{Low_Tony2014}. There the bulk model of black phosphorus is  
 \begin{equation}\label{Htb3}
 	\hat{H}_{TB}=\left(\begin{array}{cc}
 		\eta_x k_x^2+\eta_y k_y^2+\eta_z k_z^2 & \Delta/2+\gamma_x k_x^2+\gamma_y k_y^2+\gamma_z k_z^2+\mathrm{i} \chi k_x \\
 		\Delta/2+\gamma_x k_x^2+\gamma_y k_y^2+\gamma_z k_z^2-\mathrm{i} \chi k_x & \eta_x k_x^2+\eta_y k_y^2+\eta_z k_z^2
 	\end{array}\right),
 \end{equation}
which is simply the Hamiltonian \equa{Htb} with the $k_z^2$ terms added to the matrix elements, where $k_z$ is the $z$-component of the electronic momentum measured relative to the Z point. These terms simply add the $z$-direction dispersions $k_z^2/2m_{c/v,z}$ to the conduction and valence bands\cite{Low_Tony2014}, where $m_{c/v,z}=1/2(\eta_z\pm\gamma_z)$. The probe matrix elements are just quantitatively modified considering the nonzero wave vector of the probe light along the $z$-direction, which we analyze below.

We introduce another useful symmetry of the crystal structure here under the $z$-glide mirror $S_z$: $z\rightarrow -z,\, x\rightarrow x+a/2,\, y\rightarrow y+b/2$ where $a/b$ is the lattice constant along the AC/ZZ direction. With  the definition of the orbitals in \ref{fig:schematics}a, we find $\hat{S}_z\phi_A=\phi_C, \hat{S}_z\phi_B=\phi_D$, meaning $\hat{S}_z\psi_{1/2}=\psi_{1/2}$ (the basis in \equa{Htb}). 
If the probe field is in the ZZ direction and  the incident plane is the $z-y$ plane, its vector potential is $\left(A^y_{\text{probe}}\vec{e}_y+A^z_{\text{probe}}\vec{e}_z\right)\exp\left[{-\mathrm{i}\left(\omega_{\text{probe}}t-q_yy-q_zz\right)}\right]+\text{c.c.}$. 
In the zero $q_z$ limit, the probe matrix element $\langle f |  \hat{j}_z |\psi_{1/2} \rangle $ due to $A^z_{\text{probe}}$ is zero because of the even parities of $\psi_{1/2}$  and $f$ and the odd parity of $\hat{j}_z$ under $S_z$.
As before, linear expansion in $q_z$ gives a nonzero quadruple probe matrix element: $\langle f | q_z \hat{z} \hat{j}_z | \psi_{1/2} \rangle \equiv b_1^z/\sqrt{2}$.  
Applied to the conduction and valence band wave functions around the Z point,  we get $M_{fc} \approx b_1^y+b_1^z$ and $M_{fv} \approx -(b_1^y +b_1^z) i\chi k_x/G$. The matrix elements for all the probe configurations are summarized in Supplementary Table 2. For the cases relevant to our experiment ($z-y$ incident plane), it is obvious that Eqs.~\ref{eqn:amplitude} and  \ref{eqn:t1t2} for the ARPES amplitude remain the same except for an overall factor.

\subsection{Data availability}
All data are processed by lgor Pro 9.05 software. All data needed to evaluate the conclusions in the paper are available within the article and its Supplementary Information files. All data generated during the current study are available from the corresponding author upon request.

\subsection{Code availability}
The codes used for the calculations in this study are available from the corresponding author upon request.



\section*{References}

\begin{addendum}
	\item[Acknowledgements] We thank Ren-Bao Liu for helpful discussions. This work is supported by the National Natural Science Foundation of China (Grant Nos.~52388201, 12234011, 12374291, 12421004, 92250305, and 52025024), National Key R\&D Program of China (Grant Nos.~2021YFA1400100 and 2020YFA0308800). C.B. acknowledges support from the Project funded by China Science Foundation (Grant No. BX20230187) and the Shuimu Tsinghua Scholar Program. M.S. acknowledges support from SNSF Ambizione Grant No.~PZ00P2-193527.
	
	\item[Author Contributions] S.Z. conceived the research project. C.B., H.Zhong and T.L. performed the TrARPES measurements and analyzed the data. X.C., T.S., X.T., H.Zhang, P.Y. and W.D. contributed to the data analysis and discussions.  F.W. and H.Zhong grew the black phosphorus single crystal. M.S. performed the first-principles calculation. T.X. and Z.S. provided the two-band model analysis. C.B. and S.Z. wrote the manuscript, and all authors contributed to the discussions and commented on the manuscript.

	\item[Competing Interests] The authors declare that they have no competing interests.

\end{addendum}

\end{document}